\input{aipcheck}

\documentclass[
    ,final            
  ]
  {aipproc}

\layoutstyle{8x11single}
\usepackage{indentfirst}

\begin{document}

\title{Quark scalar, axial and tensor charges in the Schwinger-Dyson formalism}




\author{Nodoka Yamanaka}{
  address={2-1 Hirosawa, Wako, 351-0198 Saitama, Japan}
}

\begin{abstract}
The quark scalar, axial and tensor charges of nucleon are calculated in the Schwinger-Dyson formalism.
We first calculate these charges in the rainbow-ladder truncation using the IR cut quark-gluon vertex, and show that the result is in agreement with the known data.
We then perform the same calculation with the phenomenological IR singular quark-gluon vertex.
In this case, the Schwinger-Dyson equation does not converge.
We show that this result suggests the requirement of additional corrections to the rainbow-ladder truncation, due to the interaction between quark and gluons in the deep IR region.
\end{abstract}

\maketitle


\section{Introduction}

The quark charges of nucleon provide useful informations on the nucleon structure and on the nonperturbative physics of quantum chromodynamics (QCD).
Among them, the quark scalar, axial and tensor charges are very attractive observables, and many studies were done so far \cite{adler,aidala}.

The quark scalar charge of nucleon $\left< N | \, \bar q q | N \right>$ probes the relativistic quarks in the nucleon, since the difference of the scalar charge with the quark vector charge $\left< N | \, \bar q \gamma^0 q | N \right>$ gives the relativistic lower components of the Dirac matrix (in the Dirac representation).
This quantity is also interesting in many other contexts. 
It gives the current quark mass contribution to the nucleon mass, and also the interaction between nucleon and the dark matter in many models, such as the supersymmetry \cite{susydm}.
The quark scalar charge is studied through the pion-nucleon sigma-term $\sigma_{\pi N } \equiv m_q \left< N | \, \bar q q | N \right>$, and recent extractions from experimental data and lattice QCD simulations both point values in the range of 40-60 MeV \cite{sigmatermexp,sigmatermlattice,green,bhattacharya,ren}.

The axial charges of nucleon $\left< N (p,s) | \, \bar q \gamma^\mu \gamma_5 q | N (p,s) \right> = \Delta q\, s^\mu $ is sensitive to the longitudinally polarized quark in the longitudinally polarized nucleon \cite{axialjaffe}.
In the nonrelativistic quark model, the prediction of the total quark spin (axial charge) is $\Delta \Sigma \equiv \sum_q \Delta q = 1$, but the experimental value is much smaller \cite{compass}:
\begin{equation}
\Delta \Sigma 
= 0.32 \pm 0.03 \pm 0.03
.
\end{equation}
This result poses the unresolved problem of ``proton spin crisis".
The lattice QCD gives larger results $\Delta \Sigma \sim 0.6$ \cite{bhattacharya}.
For the isovector axial charge of the proton, the experimental data \cite{ucna}
\begin{equation}
g_A \equiv \Delta u - \Delta d
=
1.27590 \pm 0.00239 _{-0.00331}^{+0.00377}
,
\end{equation}
is also smaller than the nonrelativistic quark model prediction $g_A = \frac{5}{3}$.
For $g_A$, the lattice QCD studies give consistent results \cite{yaoki,axiallattice}.

The tensor charge of nucleon $\left< N (p,s) | \, \bar q i\sigma^{\mu \nu} \gamma_5 q | N (p,s) \right> = 2 \delta q (s^\mu p^\nu -s^\nu p^\mu )$ probes the transversely polarized quarks in the transversely polarized nucleon \cite{chiralodd}.
This quantity, like the axial charge, also gives the quark spin contribution to the nucleon spin in the nonrelativistic limit.
The quark tensor charge, combined with the axial charge, serves to probe the relativistic structure of the polarized quarks inside the nucleon, in the same way as the scalar and vector charges.
The tensor charge also has an important role in particle physics, since it gives the contribution of the quark electric dipole moment to the nucleon electric dipole moment, a sensitive probe of CP violation beyond the standard model \cite{edmreview}.
This quantity is chiral-odd, thus more difficult to study experimentally than the axial charge, and it was so far studied in many models \cite{adler,tensormodel,tensorsde,robertstensor}.
The recent extraction of the proton tensor charge from the experimental data of semi-inclusive deep inelastic scattering gives (at the renormalization point $\mu = 1$ GeV) \cite{bacchetta}
\begin{eqnarray}
\delta u 
&=& 
0.57 \pm 0.21
,
\\
\delta d 
&=& 
-0.18 \pm 0.33
.
\end{eqnarray}
Here we can also see that the total quark tensor charge $\sum_q \delta q \sim 0.5$ is less than 1, the nonrelativistic quark model prediction.
The lattice QCD studies of the quark tensor charge of nucleon also yield close results \cite{green,bhattacharya,yaoki,saoki}.

As seen above, the quark charges of nucleon have extensively been studied experimentally, using models, and from lattice QCD.
Their physical picture from the point-of-view of the quarks and gluons, however, is still not clear, and we need some analytical approach based on QCD to unveil the nonperturbative physics.
As a way to study analytically the nonperturbative physics of QCD, we have the Schwinger-Dyson (SD) formalism \cite{alkofer}.
In this work, we study the quark scalar, axial, and tensor charges in the SD formalism.
In the next section, we give the result of the SD equations for the above charges using the IR cut quark-gluon vertex.
In the subsequent section, we then do the same analysis with the IR singular quark-gluon vertex and discuss the reason of the failure.
The last section is devoted to the summary.

\section{\label{sec:ircut}Analysis with the IR cut quark-gluon vertex}

\subsection{Setup of the Schwinger-Dyson formalism}

To discuss the nonperturbative effect of the nucleon charges in the SD formalism, the Faddeev equation must be solved \cite{faddeev}.
However, in this study, we evaluate the contribution of the single quark to the nucleon charges as a first step.
The gluon dressing effect to the quark scalar, axial, and tensor charges is calculated in the the rainbow-ladder truncation \cite{tensorsde,scalarsde}.
The many-body effect will be studied by combining the obtained single quark charge \`{a} la nonrelativistic quark model.

\begin{figure}[htb]
\includegraphics[width=6.1cm,angle=-90]{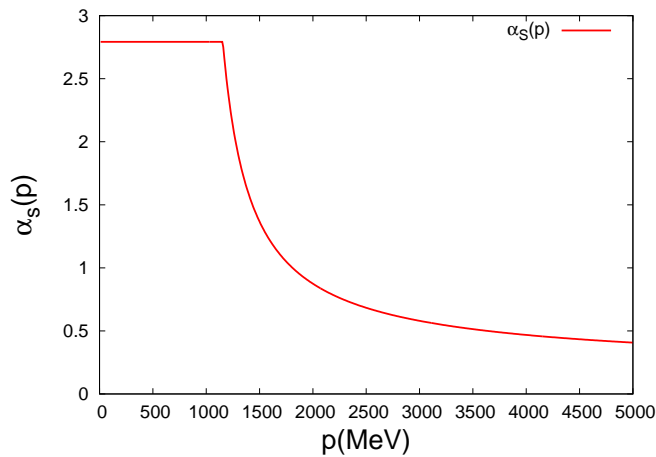}
\caption{\label{fig:alpha_s}
The effective product of the gluon dressing function and the quark-gluon vertex used in the IR cut analysis.
}
\end{figure}

We now describe the quark propagator used in the IR cut analysis.
The quark propagator is solved in the rainbow-ladder truncation, with the following Ansatz for the product of the quark-gluon vertices and the gluon dressing function
\begin{equation}
\frac{g_s^2}{4\pi} Z_g (q^2) \gamma^\mu \times \Gamma^\nu (q , k ) 
\rightarrow
\alpha_s (q) \gamma^\mu \times \gamma^\nu
\, ,
\label{eq:rlapprox}
\end{equation}
where $Z_g (q^2)$ is the gluon dressing function, and $\Gamma^\nu (q,k) $ is the dressed quark-gluon vertex \cite{ballchiu,qgvertex,fischer}.
Here the effective product of the gluon dressing function and the quark-gluon-vertex is given by the simple Ansatz \cite{higashijima}
\begin{eqnarray}
\alpha_s (p) =
\left\{
\begin{array}{ll}
\frac{24\pi}{11N_c - 2N_f} & (p<p_{\rm IR}) \cr
\frac{12\pi}{11N_c - 2N_f} \frac{1}{\ln (p^2 / \Lambda_{\rm QCD}^2)} & (p\geq p_{\rm IR}) \cr
\end{array}
\right.
\, ,
\end{eqnarray}
where $N_c = N_f =3 $, and $p_{\rm IR}$ satisfies $\ln (p_{\rm IR}^2 / \Lambda_{\rm QCD}^2)= \frac{1}{2}$, with $\Lambda_{\rm QCD}$ = 900 MeV.
The shape of $\alpha_s (p)$ is plotted in Fig. \ref{fig:alpha_s}.
In this setup, the resulting chiral condensate and the pion decay constant are $-(238\, {\rm MeV})^3$ and $f_\pi = 70$ MeV, respectively.

\begin{figure}[htb]
\includegraphics[width=12cm]{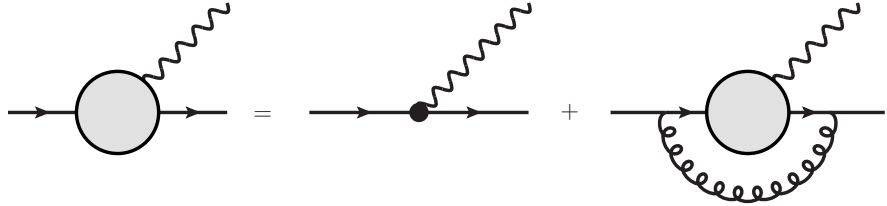}
\caption{\label{fig:SDE}
The SD equation for the gluon dressing effect to the quark scalar, axial, and tensor charges of nucleon.
The quark-gluon vertices, the quark and gluon propagators are dressed.
}
\end{figure}

We now solve the quark charge SD equation.
The self-consistent equation to be solved is shown in Fig. \ref{fig:SDE}.
We again use the rainbow-ladder truncation, with the derived dynamical quark propagator together with two dressed quark-gluon vertices and the dressed gluon propagator as inputs.
In this section, we use the same effective product (\ref{eq:rlapprox}) to replace two quark-gluon vertices and the gluon propagator.
This Ansatz is not consistent with that used for the quark propagator, since Eq. (\ref{eq:rlapprox}) was used to replace the product between the gluon propagator and only one quark-gluon vertex.
The difference, however, should only be relevant in the IR region, since the UV behavior is controlled by the same gluon dressing function.
As the quarks and gluons are confined in the nucleon with the radius 0.5 fm, the deep IR contribution should not be important, and small changes of the quark-gluon vertex in the IR region will not change the qualitative result.
In the vertex SD equation of the IR cut analysis, we therefore use the same effective expression of Eq. (\ref{eq:rlapprox}) to replace the gluonic contribution.

\subsection{Result}

The result of the IR cut analysis for the quark scalar, axial, and tensor charges of nucleon obtained after solving the vertex SD equations in the rainbow-ladder truncation is shown in Tables \ref{table:isoscalar} and \ref{table:isovector}.
We see a reasonable agreement between our result and the experimental or lattice QCD data\footnote{While preparing this proceeding, Ref. \cite{robertstensor} appeared, where the tensor charge of nucleon was solved with the Faddeev equation. The result is quite consistent with the values obtained in our simple setup.}.

The scalar charge is enhanced by the gluon dressing effect.
This is because the scalar charge is not conserved by the Ward identity, and becomes larger as the quark takes a long world line in the 4 dimensional space-time.
The exchanges of gluons lengthen the world line of the quark or create disconnected quark loops, so that the relativistic effect enhances the quark scalar charge.

The axial and tensor charges are suppressed by the gluon dressing effect.
This is due to the superposition of the spin flipped states of the quark, as the gluon carries spin 1 and its emission/absorption flips the quark polarization.

\begin{table}
\begin{tabular}{lccc}
\hline
& Experiment & Lattice & Our result \\
\hline
Scalar & 10 & 10 - 15 & 27 \\
Axial & 0.32 & 0.6 & 0.85 \\
Tensor & 0.5 & 0.6 & 0.6 \\
\hline
\end{tabular}
\caption{Isoscalar proton charges.
}
\label{table:isoscalar}
\end{table}

\begin{table}
\begin{tabular}{lccc}
\hline
& Experiment & Lattice & Our result \\
\hline
Scalar & 1 & 1 & 9 \\
Axial & 1.27 & 1.2 & 1.4 \\
Tensor & 1.0 & 1 & 1.0 \\
\hline
\end{tabular}
\caption{Isovector proton charges.
}
\label{table:isovector}
\end{table}

\section{\label{sec:singular}Analysis with the IR singular quark-gluon vertex: failure}

\subsection{Setup of the Schwinger-Dyson formalism}

\begin{figure}
  \includegraphics[width=6.1cm,angle=-90]{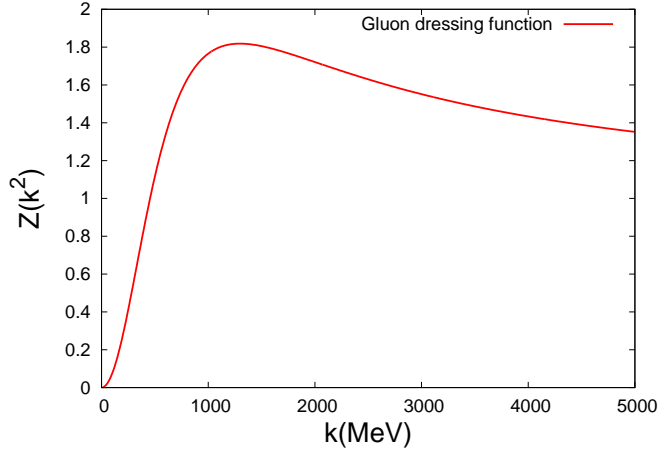}
  \caption{\label{fig:gluon_dressing}The gluon dressing function.}
\end{figure}

\begin{figure}
  \includegraphics[width=6.1cm,angle=-90]{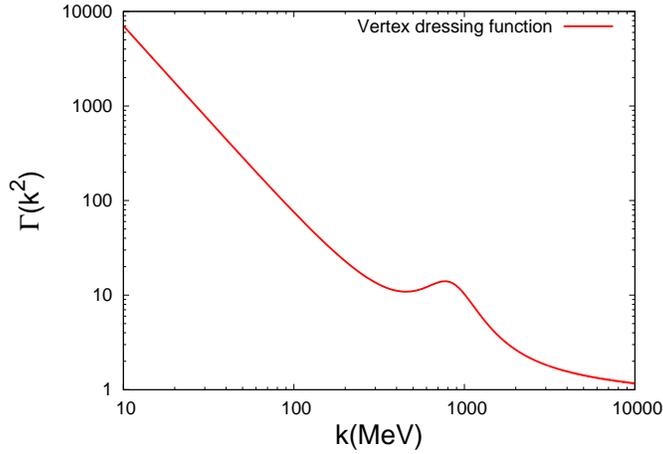}
  \caption{\label{fig:vertex_dressing}IR singular vertex dressing function, plotted in the logarithmic scale.}
\end{figure}

The next step is to consider the IR singular quark-gluon vertex.
The IR singular quark-gluon vertex can phenomenologically explain the large $\eta'$ mass \cite{singularqgvertex}.
In this section, we use the same quark-gluon vertex and gluon propagator as in Ref. \cite{singularqgvertex}, with $\kappa =0.5$.
This choice gives a IR finite gluon propagator ($\propto k^2$ gluon dressing function in the IR limit $k\rightarrow 0$), and a quadratically divergent quark-gluon vertex ($\frac{1}{k^2}$ for $k \rightarrow 0$), to respect the IR behavior of the gluon propagator suggested by recent lattice calculations in the Landau gauge \cite{irgluonlattice,gluonpropagatorsde}.
The gluon dressing function and the quark-gluon vertex are plotted in Fig. \ref{fig:gluon_dressing} and \ref{fig:vertex_dressing}, respectively.
With this setup, the calculated pion decay constant is $f_\pi = 92$ MeV.

\subsection{Result}

\begin{figure}
\includegraphics[height=.2\textheight]{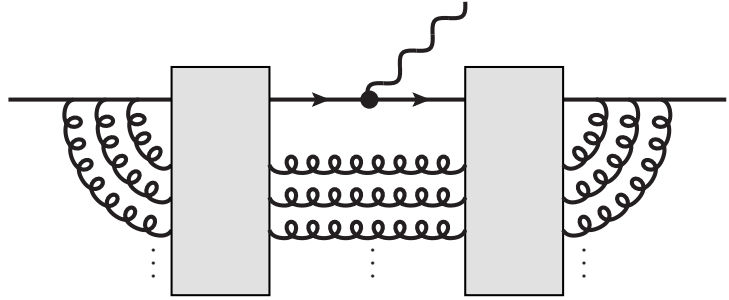}
\caption{\label{fig:beyond_rainbow}
Correction beyond rainbow-ladder approximation in the vertex SD equation.
}
\end{figure}

We have tried to calculate the quark charge SD equation with the same procedure as for the IR cut analysis.
However, it was not possible to converge the SD equation.
The SD equation has not converged due to the large IR contribution to the momentum integral.
Although being phenomenological, the IR singular quark-gluon vertex should better describe the IR feature than the IR cut vertex.
This failure thus suggests the existence of additional important contribution beyond the rainbow-ladder approximation \cite{fischer,beyondrainbow} in the deep IR region.
The solution we suggest is the following.
The rainbow-ladder approximation in the vertex SD equation actually neglects the interaction among gluons in the intermediate states.
As the quarks and gluons are strongly interacting in the IR, we actually need the N-gluon+quark interacting kernel which is expected to damp the deep IR contribution of the rainbow-ladder process.
The schematic diagram of this process is drawn in Fig. \ref{fig:beyond_rainbow}.
We expect that this IR suppression is equivalent to an effective setup which resembles the IR cut analysis given in the previous section.

\section{Summary}

In this work, we have calculated the gluon dressing effect to the quark scalar, axial and tensor charges in the Schwinger-Dyson formalism with a simple setup.
We have first discussed the case with IR cut quark-gluon vertex, for which the IR contribution is damped.
As a result, the quark scalar charge is enhanced due to nonrelativistic effect.
The axial and tensor charges are, in contrast, suppressed since the dynamical charges are given by the superposition of the spin flipped intermediate quark states due to the gluon emission and absorption.
This result is in qualitative agreement with the known experimental and lattice QCD data.
The inclusion of the IR singular quark-gluon vertex however fails in the rainbow-ladder approximation.
This is because we have not taken into account the long range interaction between each gluon and quark in the intermediate state.
A priori, strong damping of the IR contribution is expected due to the confinement.
To resolve this problem, the study of effects beyond the rainbow-ladder approximation is required \cite{beyondrainbow}.
Moreover, our simple setup based on the nonrelativistic quark model should be improved by considering the many-body effect via relativistic Faddeev formulation \cite{faddeev}.


\begin{theacknowledgments}
The author thanks Xiu-Lei Ren for useful discussion and comments.
This work is supported by the RIKEN iTHES project.
\end{theacknowledgments}



\bibliographystyle{aipproc}   

\bibliography{sample}

\IfFileExists{\jobname.bbl}{}
 {\typeout{}
  \typeout{******************************************}
  \typeout{** Please run "bibtex \jobname" to optain}
  \typeout{** the bibliography and then re-run LaTeX}
  \typeout{** twice to fix the references!}
  \typeout{******************************************}
  \typeout{}
 }


\end{document}